\documentstyle[preprint,pra,aps]{revtex}
\begin{document}
\draft
\title{\bf Electric dipole moments of
Hg, Xe, Rn, Ra, Pu, and TlF induced by the nuclear Schiff moment
and limits on time-reversal violating interactions}
\author{V.A. Dzuba, V.V. Flambaum,
and J.S.M. Ginges}
\address{School of Physics, University of New South Wales,
Sydney 2052, Australia}
\author{M.G. Kozlov}
\address{Petersburg Nuclear Physics Institute,
Gatchina 188300, Russia}
\date{\today}
\maketitle

\tightenlines
\begin{abstract}

We have calculated the atomic electric dipole moments (EDMs)
induced in $^{199}$Hg, $^{129}$Xe, $^{223}$Rn, $^{225}$Ra,
and $^{239}$Pu by their respective nuclear Schiff moments $S$.
The results are (in units $10^{-17}S(e~{\rm fm}^{3})^{-1}e~{\rm cm}$):
$d(^{199}{\rm Hg})=-2.8$,
$d(^{129}{\rm Xe})=0.38$,
$d(^{223}{\rm Rn})=3.3$,
$d(^{225}{\rm Ra})=-8.5$,
$d(^{239}{\rm Pu})=-11$.
We have also calculated corrections to the
parity- and time-invariance-violating ($P,T$-odd)
spin-axis interaction constant in TlF.
These results are important for the interpretation of atomic
and molecular experiments on EDMs in terms of fundamental
$P,T$-odd parameters.

\end{abstract}
\vspace{1cm}
\section{Introduction}

Recently, a stringent upper limit on the ground-state atomic
electric dipole moment (EDM) of $^{199}$Hg was obtained
\cite{romalis01},
\begin{equation}
\label{eq:Hglimit}
d(^{199}{\rm Hg})=-(1.06\pm 0.49\pm 0.40)\times 10^{-28}~e~{\rm cm}~ .
\end{equation}
(The respective errors are statistical and systematic.)
This is the best experimental upper limit on an atomic EDM to date.
Combined with calculations, this limit can be interpreted in
terms of limits on fundamental parity- and time-invariance-violating
($P,T$-odd) parameters.
These limits tightly constrain competing theories of $CP$ violation.

Hg has closed electronic subshells, $J=0$.
The measurement (\ref{eq:Hglimit}) is therefore sensitive to
$P,T$-violating mechanisms which originate from the nucleus.
The $P,T$-odd nuclear moment that can induce a Hg EDM is
the Schiff moment (the nuclear EDM is screened by atomic electrons
\cite{schiff63} and the magnetic quadrupole moment does
not contribute due to zero electron angular momentum).
Note that there are other mechanisms by which the Hg
EDM can be induced, such as the $P,T$-odd electron-nucleon
interaction (see, e.g. \cite{khriplovich}),
and the electron EDM (it contributes due to the hyperfine
interaction) \cite{fortson83,flambaum85}.

A value for the EDM of Hg induced by the Schiff moment $S$
was estimated in \cite{flambaum85i},
$d(^{199}{\rm Hg})=-4\times 10^{-17}S
(e~{\rm fm}^{3})^{-1}e~{\rm cm}$~.
This value has been used for the interpretation of the
measurements of the Hg EDM in terms of $P,T$-odd nuclear parameters.
However, this value was obtained indirectly
from an atomic calculation \cite{mp} of the EDM of Hg
induced by the $P,T$-odd electron-nucleon tensor interaction.

In this work we have performed numerical calculations of the EDMs
induced by $S$ in Hg and in other diamagnetic atoms of current
experimental interest.
Our result for Hg,
\begin{equation}
d(^{199}{\rm Hg})=-2.8\times 10^{-17}\Big(
\frac{S}{e~{\rm fm}^{3}} \Big) e~{\rm cm} \ ,
\end{equation}
differs from the previous estimate (about $40\%$ smaller) and
places a more conservative constraint on $S$,
and hence the underlying $P,T$-odd mechanisms that induce it.

The other atoms we have studied in this work are $^{129}$Xe, $^{223}$Rn,
$^{225}$Ra, and $^{239}$Pu, in their ground-states.
There has been a recent measurement
of the EDM induced in $^{129}$Xe \cite{rosenberry01},
\begin{equation}
\label{eq:Xelimit}
d(^{129}{\rm Xe})=(0.7\pm 3.3\pm 0.1)\times 10^{-27}e~{\rm cm} \ .
\end{equation}
A calculation for the Xe EDM induced by $S$ has previously been
performed at the Hartree-Fock level \cite{dzuba85},
$d(^{129}{\rm Xe})=0.27\times 10^{-17}S
(e~{\rm fm}^{3})^{-1}e~{\rm cm}$~.
In this work we have found that with core polarization included this
value becomes
\begin{equation}
d(^{129}{\rm Xe})=0.38\times 10^{-17}\Big(
\frac{S}{e~{\rm fm}^{3}} \Big) e~{\rm cm} \ .
\end{equation}
This value is $40\%$ larger than the previous calculation
and hence places a tighter constraint on the $P,T$-odd parameters
extracted from the Xe measurement.

No experiments or direct calculations have been performed
for the ground-state EDMs of Rn, Ra, and Pu.
In these atoms there is a possibility for
an enhanced Schiff moment due to static \cite{spevak97} or even
soft \cite{engel00} octupole deformation.

Measurements of EDMs of Rn isotopes are in preparation \cite{chupp}.
Rn is a heavier analog of Xe.
As well as the nuclear enhancement in Rn compared to Xe,
there is an electronic enhancement arising from the increase
of the electronic density near the nucleus with higher $Z$.
An estimate of the Rn EDM was made in Ref. \cite{spevak97}
by extrapolation of the EDM of Xe \cite{dzuba85}, taking
into account the enhancement due to increase in $Z$.

Also, estimates for Ra \cite{spevak97} and Pu \cite{flambaum02}
have previously been carried out using the same extrapolation method
as for Rn, although here the results were extrapolated from the
estimate of the EDM of Hg \cite{flambaum85i}.

In Section \ref{s:con} (Table \ref{tab:finaledm}) the values of previous
calculations/estimates for the atomic EDMs induced by $S$ in
Xe, Rn, Ra, Hg, and Pu and the results of this work are
presented.

The experimental study of the $P,T$-odd effects induced by the nuclear Schiff
moments is not restricted to atoms. In the paper \cite{tlfedm} the
$P,T$-odd spin-axis interaction in TlF molecule was measured.
From this experiment a limit can be placed
on the Schiff moment of the $^{205}$Tl nucleus. Recently the accuracy of the
molecular calculations was significantly increased (see \cite{petrov02} and
references therein) and here we want to use these new molecular calculations
to obtain a more reliable limit on $S({^{205}{\rm Tl}})$.
This is dealt with in Section \ref{s:tlf}.

\section{Method of calculation}

The nuclear Schiff moment $S$ produces a $P,T$-odd
electrostatic potential $\varphi$.
The interaction Hamiltonian
\begin{equation}
H_{W}=\sum _{i}h_{W}^{i}=-e\sum _{i}\varphi ({\bf R}_{i})
\end{equation}
mixes states of opposite parity and induces a static atomic EDM.
In previous calculations the contact form for the
electrostatic potential was used,
\begin{equation}
\label{eq:phi0}
\varphi ({\bf R})=4\pi {\bf S}\cdot
\mbox{\boldmath$\nabla$}\delta ({\bf R}) \ .
\end{equation}
However, for relativistic electrons the matrix element of this
potential diverges.
Usually this problem is solved by a cutoff of the electron
wave functions at the nuclear surface.
In the work \cite{flambaum02} it was found that there is
a more convenient form for $\varphi$ suitable for
relativistic atomic calculations,
\begin{equation}
\label{eq:phi}
\varphi ({\bf R})=-\frac{3 {\bf S}\cdot {\bf R}}{B}\rho (R)\ ,
\end{equation}
where $B=\int \rho (R)R^{4}dR$,
and $\rho (R)$ is the nuclear density.
This expression arises because the Schiff moment
produces a constant electric field, along the nuclear spin,
inside the nucleus \cite{flambaum02}.

The atomic EDM induced in the many-body state $N$ by the
$P,T$-odd interaction $H_{W}$ can be expressed as
\begin{equation}
\label{eq:fulledm}
d=2\sum _{M}\frac{\langle N|H_{W}|M\rangle
\langle M|D_{z}|N\rangle}{E_{N}-E_{M}} \ ,
\end{equation}
where the sum $M$ runs over a complete set of many-body states,
$E_{N}$ and $E_{M}$ are atomic energies, and
$D_{z}$ is the atomic electric dipole operator.

The atoms we have studied in this work are $^{129}$Xe, $^{223}$Rn,
$^{225}$Ra, $^{199}$Hg, and $^{239}$Pu.
Of course, the method of calculation of the electronic states
depends on the electron configuration of the atom.
The starting point for the calculations is to self-consistently
solve the single-particle relativistic Hartree-Fock (HF) equation
\begin{eqnarray}
\label{eq:sch}
&&h_{0}|i\rangle =\epsilon _{i} |i\rangle \\
\label{eq:hf}
&&h_{0}=c\mbox{\boldmath$\alpha$}\cdot {\bf p}
+(\beta-1)c^{2}-Z\alpha/r+V
\end{eqnarray}
for each electron $|i\rangle$ contributing to the potential $V$.

For the noble gases Xe and Rn the calculations
are performed in the Hartree-Fock approximation in the
$V^{N}$ potential. This corresponds to solving (\ref{eq:sch},\ref{eq:hf})
for the $N$ electrons of the atom, that is, in a self-consistent
potential $V^{N}$ formed from all $N$ electrons.

The atoms Ra and Hg can be treated as closed shell systems or as
atoms with two electrons above closed shells.
Correspondingly, we have performed two
separate calculations for these atoms: one in the $V^{N}$ potential; and
a more detailed calculation in the $V^{N-2}$ potential
(solving (\ref{eq:sch},\ref{eq:hf}) for the $N-2$ electrons in the atom)
with correlations between the external electrons and between
the external electrons and the core included.
The interaction of the two
external electrons is treated using the configuration interaction
(CI) method and the correlations of the external electrons with the core
accounted for by many-body perturbation theory (MBPT).
This technique, the CI+MBPT method,
was developed in the works \cite{dzuba96,dzuba98}.

Pu corresponds to an open shell atom, with electron
configuration $5f^{6}7s^{2}$.
We have performed a simple calculation for this atom,
in the $V^{N}$ approximation, but with the contribution
of the open $f$-shell weighted according to its occupancy.
(This simple calculation in the $V^{N}$ potential, with no correlations
accounted for, is justified by comparison of the
results of the calculations for Hg and Ra in the potentials $V^{N}$
and $V^{N-2}$; see Section \ref{s:hgra}.)

As a test of our wave functions we have performed calculations of the
ionization potentials and the scalar polarizabilities $\alpha $
of the ground-state for each atom.
These can then be compared with the available experimental data.
It is easy to calculate the polarizability
\begin{equation}
\alpha =-2\sum _{M}
\frac{|\langle N|D_{z}|M\rangle |^{2}}{E_{N}-E_{M}} \ ,
\end{equation}
by replacing the operator $H_{W}$ in
(\ref{eq:fulledm}) by the dipole operator $D_{z}$.

Below we outline the $V^{N}$ and $V^{N-2}$ calculations.

\subsection{$V^{N}$ approximation}

In the $V^{N}$ calculation, we can write the atomic EDM induced
by the Schiff moment as
\begin{equation}
\label{eq:edmn}
d=2\sum _{n}\langle \delta n_{W}|d_{z}|n\rangle \ ,
\end{equation}
where the sum runs over the core states $|n\rangle$,
$d_{z}$ is the single-particle dipole operator,
and $|\delta n _{W}\rangle$ denotes
the correction to the state $|n\rangle$ due to the $P,T$-odd
Hamiltonian $h_{W}$.
The correction $|\delta n _{W}\rangle$ can be expressed as
\begin{equation}
\label{eq:deltansum}
|\delta n _{W}\rangle =
\sum _{\alpha}\frac{\langle \alpha |h_{W}|n\rangle}
{\epsilon _{n}-\epsilon _{\alpha}}|\alpha\rangle \ ,
\end{equation}
where $|\alpha\rangle$ corresponds to an excited state.
It is found by solving the equation
\begin{equation}
\label{eq:deltan}
(h_{0}-\epsilon _{n})|\delta n_{W}\rangle
= -h_{W}|n\rangle \ .
\end{equation}
(Of course, calculating the correction to $|n\rangle$ due to the
electric dipole (E1) field and taking the matrix element of the
weak Hamiltonian is equivalent.)

To include polarization of the core due to the fields
$h_{W}$ and $d_{z}$, we need only include
the polarization due to one field, e.g. by replacing
$h_{W}$ in (\ref{eq:deltan}) by
$h_{\tilde{W}}=h_{W}+\delta V_{W}$.
(The correction $\delta V_{W}$ accounts for the change
in the self-consistent potential $V^{N}$ due to the
modification of the wave functions. For more on the
time-dependent Hartree-Fock (TDHF) method, or random-phase
approximation (RPA) with exchange, see, e.g.,
\cite{dzuba87}.)
This is because all the states $|n\rangle$ belong to the core
(that is, they are included in the Hartree-Fock potential),
so
$\sum_{n}\langle \delta n_{\tilde{W}}|d_{z}|n\rangle
=\sum _{n}\langle \delta n_{W}|
d_{z}+\delta V_{d}|n\rangle$.

The scalar polarizabilities are easily calculated by replacing the
correction $|\delta n_{W}\rangle$ due to the $P,T$-odd field by
the correction $|\delta n_{d}\rangle$ due the E1 field.

\subsection{$V^{N-2}$ approximation}

Hg and Ra can be considered as atoms with two valence electrons above closed
cores, $[1s...5d^{10}]$ and $[1s...5d^{10}6s^{2}6p^{6}]$, respectively. The
calculations performed in the $V^{N-2}$ approximation correspond to the
CI+MBPT method developed in the works \cite{dzuba96,dzuba98}. An effective
Hamiltonian is constructed for the valence electrons which is formed using
MBPT for the interaction of the valence electrons with the core. In this way
the correlations between the valence electrons $6s^{2}$ (for Hg) and $7s^{2}$
(Ra) are treated using the CI method, while the valence-core correlations are
treated using MBPT. (For more on this procedure, please refer to
\cite{dzuba96,dzuba98}.)

Note that calculations in the CI+MBPT method can be performed in a different
potential, even in $V^{N}$. The final accuracies for $V^{N-2}$ and $V^{N}$
potentials are comparable; however, calculations in $V^{N-2}$ for Hg and Ra
are somewhat simpler --- there are no subtraction diagrams, see, e.g., Ref.
\cite{dzuba96}.

For Hg the CI states were formed from the one-electron basis sets
$1s-12s$, $2p-13p$, $3d-14d$, and $4f-15f$.
The core $1s-5d$ and the states $6s$, $6p$, $6d$, $5f$ were
HF ones, while the rest were virtual orbitals.
The virtual orbitals were formed using the recurrent
procedure, similar to that used in \cite{bogdanovich83}.
Full CI was made for the two valence electrons.
Two sets of virtual orbitals were constructed to test
the saturation of the CI space. The results appeared to be very close, so we
concluded that saturation was reached.

MBPT calculations generally require more basic functions for high energies
and more partial waves,
so the number of virtual orbitals was increased to construct the
MBPT basis set: it included
$1s-21s$, $2p-22p$, $3d-23d$, $4f-21f$, $5g-18g$, and $6h-19h$.
For Ra the CI basis set was somewhat larger and included orbitals
$1s-15s$, $2p-16p$, $3d-17d$, and $4f-17f$.
The extended basis set for MBPT was similar to that used for Hg.

\section{Results}

\subsection{Xenon and radon}

The binding energies of the ground state of
Xe and Rn performed in the HF approximation in the $V^{N}$
potential are presented in Table \ref{tab:ip}.
It is seen that the Hartree-Fock calculation serves as
a good approximation for the noble gases,
with the calculated and experimental ionization potentials
in disagreement only at the level of $1\%$.

The ground-state scalar polarizabilities $\alpha $ are listed in
Table \ref{tab:alphavn}.
Core polarization increases $\alpha $ only very slightly
(about $1\%$) from the Hartree-Fock value.
The value for Xe is in perfect agreement with the experimental
value.

Our results for the electric dipole moments induced in
Xe and Rn due to the nuclear Schiff moment are
presented in Table \ref{tab:dvn}.
The effect of core polarization
increases the value of the EDMs in Xe and Rn by about $30\%$.
Due to the increase in $Z$, the EDM induced in Rn is ten times
larger that that induced in its lighter analog Xe.

A calculation for the Xe EDM has previously been performed
at the HF level, with the result
$d(^{129}{\rm Xe})=0.27\times 10^{-17}S
(e~{\rm fm}^{3})^{-1}e~{\rm cm}$ \cite{dzuba85}.
Our HF result is in agreement with this value.

No direct atomic calculation has previously been performed
for the atomic EDM of Rn,
or for any of the remaining atoms in this work.
The estimate for Rn, based on extrapolation of the HF result
from Xe, is
$(^{223}{\rm Rn})=2.0\times 10^{-17}S
(e~{\rm fm}^{3})^{-1}e~{\rm cm}$ \cite{spevak97}.
This value is not significantly different from our HF value.

\subsection{Mercury and radium}
\label{s:hgra}

The ionization potentials of Hg and Ra performed in the $V^{N}$
approximation are listed in Table \ref{tab:ip} alongside the
results for the noble gases. The deviation from experiment
for Hg and Ra is $\sim -14\%$, indicating the importance of correlations.

The ground-state polarizabilities for Hg and Ra
performed in the $V^{N}$ approximation are listed in
Table \ref{tab:alphavn}.
We see here that core polarization increases $\alpha $,
from the HF value, by about 10\% for Hg. The effect of
core polarization on Ra is more significant, with an increase of 45\%.
For both Hg and Ra, essentially the entire HF and TDHF results for
$\alpha$ arise from the $6s^{2}$ and $7s^{2}$ contributions, respectively.
We see that the result for Hg is strongly
overestimated (about $30\%$ higher than the experimental value).
This indicates that the $V^{N}$ calculations produce
wave functions that are very poor at large distances from the nucleus
for systems with two electrons above closed shells.

Our results for the EDMs induced in Hg and Ra due to the
nuclear Schiff moment, in the $V^{N}$ approximation,
are presented in Table \ref{tab:dvn}.
The effects of core polarization in Hg and Ra are dramatic,
with the EDM increased by a factor of $2.5$ and $4.5$,
respectively.
The dominating contributions to the EDMs of Hg and Ra
come from $6s^{2}$ and $7s^{2}$, respectively.
These contributions are larger than, and of opposite sign to,
the contributions from the core.
The instability of the results motivates us to study these
atoms more carefully, taking into account valence-valence
and valence-core correlations.

The $V^{N-2}$ approximation is more appropriate for calculations
of two-electron atoms.
The results of the calculations for the binding energies of relevant
states of Hg and Ra are presented in Table \ref{tab:ipn-2}.
The calculated value for the removal energy of both $s$-electrons
for Hg and Ra deviates from experiment by about $-$10\% at the
CI level, and is then improved to +2\% with MBPT included.
This is a significant improvement compared to the $V^{N}$ result.
The accuracy of the calculations of the energies of other states
is not as good. These states are not directly relevant to
our calculations of the EDMs, however we have presented them
in the table as an indication of the accuracy of wave functions
produced at various stages of the CI+MBPT method.
At the CI stage of the calculations the energies of the
$ss$ and $sp$ states are underestimated by $10-20\%$.
Note that for $sd$ states the energies are overestimated
by about the same amount.
With MBPT corrections included, the accuracy of the
energies improves to $2-9\%$.

Results of the calculations of the ground-state
scalar polarizabilities $\alpha$ of Hg and Ra
in the $V^{N-2}$ approximation are listed in
Table \ref{tab:alphavn-2}. We see that for Hg at the
CI level of calculation $\alpha$ is strongly overestimated
(by about $60\%$).
However, both the MBPT and TDHF corrections reduce the value
so that the final result is in excellent agreement with experiment
(compare 32.99 a.u. with 34 a.u. \cite{radtsig}).
We see that for Ra the MBPT and TDHF corrections bring about
a cancellation of the CI value similar to what we see for Hg.

The results of the calculations of the EDM of Hg and Ra in the
$V^{N-2}$ approximation are presented in Table \ref{tab:dvn-2}.
We see here that for Hg the MBPT and TDHF corrections increase the CI
approximation by a factor of two.
The corrections for Ra are huge, the final value
3.6 times that of the CI value.

Even though these corrections are so large, it is certainly
interesting that the result for $d({\rm Hg})$ in the
$V^{N-2}$ approximation (with correlations) is in agreement
($\approx 10\%$) with the simple calculation in the $V^{N}$ approximation;
compare the results of Hg and Ra in
Table \ref{tab:dvn-2} with those in Table \ref{tab:dvn}.
This coincidence of the results is an argument in favour
of the stability of the final values.
Also it indicates that the simple $V^{N}$ approximation gives
accurate results for EDM calculations of closed-shell atoms.
From a consideration of the results, we take as
our final value for the EDM in Hg induced by $S$,
\begin{equation}
\label{eq:hgS}
d(^{199}{\rm Hg})=-2.8\times 10^{-17}\Big(
\frac{S}{e~{\rm fm}^{3}} \Big) e~{\rm cm} \ .
\end{equation}

We take as our final value for the EDM induced in Ra
\begin{equation}
d(^{225}{\rm Ra})=-8.5\times 10^{-17}\Big(
\frac{S}{e~{\rm fm}^{3}} \Big) e~{\rm cm} \ .
\end{equation}

We can compare these new values for Hg and Ra
with estimates calculated in Refs. \cite{flambaum85i}
and \cite{spevak97}, respectively.
The value for Hg, $d(^{199}{\rm Hg})=-4\times 10^{-17}S
(e~{\rm fm}^{3})^{-1}e~{\rm cm}$,
was obtained indirectly from an atomic calculation \cite{mp}
of the Hg EDM induced by the $P,T$-odd electron-nucleon
tensor interaction.
The radium estimate, $d(^{225}{\rm Ra})=-7\times 10^{-17}S
(e~{\rm fm}^{3})^{-1}e~{\rm cm}$, was found by
extrapolation of the Hg EDM estimate, taking account of the
enhancement due to higher atomic number $Z$.

\subsection{Plutonium}

We performed a simple calculation for Pu.
We basically used the $V^{N}$ approximation.
The unfilled $f$-shell was accounted for by
weighting the corresponding angular coefficients
according to the occupation number of the shell.
In calculations of the atomic EDM,
we can expect that Pu behaves in a similar way to
Ra and Hg (there is a closed $s$-shell, $7s^{2}$,
and an open $f$-shell).
The $f$-shell does not contribute to the EDM
at the HF level (the $f$-shell does not
penetrate the nucleus due to the centrifugal barrier,
and so does not contribute to the Schiff matrix element).
However, it can contribute due to core polarization.
Because the $V^{N}$ calculations for Ra and Hg turned
out to be sufficient, we expect that the same is true
for Pu.

Our result for the ionization potential of Pu
(see Table \ref{tab:ip}) deviates by -15\% from
experiment. The polarizability $\alpha$ is listed
in Table \ref{tab:alphavn}.
As we saw for Hg and Ra, essentially the entire
value is due to the $7s^{2}$ contribution.
The results for the EDM of Pu are presented in Table
\ref{tab:dvn}.
These values are essentially due to the contribution of the
$7s$-electrons, which are of opposite sign to the contributions
of the core and $f$-electrons. With core polarization, the contribution
of the $f$-electrons amounts only to 10\% of the final value.
The effect of core polarization increases
the HF value 4 times.
This is similar to what we see in Ra.

An estimate for the Pu EDM was performed in
Ref. \cite{flambaum02} by extrapolation from the
estimate of the Hg EDM,
$d(^{239}{\rm Pu})=-10\times 10^{-17}S
(e~{\rm fm}^{3})^{-1}e~{\rm cm}$.
Our calculation is in agreement with this estimate.

\subsection{TlF molecule}
\label{s:tlf}

All molecular calculations deal with the following matrix element:
\begin{eqnarray}
\label{tlf1}
  X &=& -\frac{2 \pi}{3}
  \langle \Psi_0|[\mbox{\boldmath$\nabla$} \cdot {\bf n},\delta({\bf R})]
  |\Psi_0 \rangle,
\end{eqnarray}
where $\Psi_0$ is the ground state wave function and $\bf n$ is the unit
vector along the molecular axis. It is clear that this matrix element is
related to the contact form of the Schiff moment potential given by
Eq.~(\ref{eq:phi0}):
\begin{eqnarray}
\label{tlf3}
  -4\pi \langle \Psi_0| {\bf S}\cdot
  [\mbox{\boldmath$\nabla$},\delta({\bf R}) ] |\Psi_0 \rangle
  &=& 6 X {\bf S}\cdot{\bf n}.
\end{eqnarray}
It is not difficult to find the correction coefficient $k_1$
which accounts for the difference between the contact form of
the Schiff moment interaction (\ref{eq:phi0}) and the more
accurate expression (\ref{eq:phi}).

The latest molecular calculation \cite{petrov02} accounted for the
correlations between valence electrons, but neglected core-valence
correlations. We saw above that the latter appears to be very important in
atomic calculations. The most important core-valence correlations in molecules
with one heavy atom are of the same nature as in atoms and can be accounted
for by an atomic calculation for the heavy atom \cite{KTMS97}.
The valence space in the calculation \cite{petrov02} included
$5s,\,5p,\,5d,\,6s$, and $6p$ electrons, leaving a rather small core
$[1s^2\dots 4f^{14}]$. For such a compact
and rigid core the dominant correlation correction is from the RPA for the
Schiff potential. We account for this correction by the coefficient $k_2$.

The final expression, which includes the two corrections discussed above,
has the form:
\begin{eqnarray}
\label{tlf4}
  \langle \Psi_0| H_W |\Psi_0 \rangle &=& 6\, k_1\, k_2\, X
  {\bf S}\cdot{\bf n}.
\end{eqnarray}
The dominant contribution to the molecular matrix
element~(\ref{tlf4}) comes from atomic matrix elements for Tl of the form
\begin{eqnarray}
\label{tlf6}
  \langle ns | H_W | mp_z \rangle
  &=& \frac{1}{3} \langle ns_{1/2} | H_W | mp_{1/2} \rangle
  +  \frac{2}{3} \langle ns_{1/2} | H_W | mp_{3/2} \rangle.
\end{eqnarray}
This allows us to approximate the correction coefficients:
$k_i \approx \frac{1}{3}k_i(s_{1/2},p_{1/2})+
\frac{2}{3}k_i(s_{1/2},p_{3/2})$. Within this method
we calculated the coefficients $k_i$ for the Tl atom and found:
\begin{eqnarray}
\label{tlf5}
  k_1 &=& 0.89; \qquad k_2\, =\, 1.10.
\end{eqnarray}
We see that both corrections are relatively small and, therefore, our simple
model calculation is justified. Note that the core here is much smaller than
in our previous atomic calculations. Because of that the RPA correction here
is only 10\%. Other core-valence correlations tend to decrease more rapidly
with the energy of the core orbitals, and therefore are less important here.
Using the result $X=7635$~{a.u.} from \cite{petrov02} and Eqs.~(\ref{tlf4})
and~(\ref{tlf5}) we get as our final answer:
\begin{eqnarray}
\label{tlf7}
  \langle \Psi_0| H_W |\Psi_0 \rangle &=& 4.47\times 10^4
  ({\bf S}\cdot {\bf n})\,\, \mbox{a.u.} \ .
\end{eqnarray}

\section{Conclusion}
\label{s:con}

The most accurate measurement of the $P,T$-odd spin-axis interaction
in the TlF molecule \cite{tlfedm}
combined with Eq.~(\ref{tlf7}) gives the limit:
\begin{equation}
\label{tlf8}
S(^{205}{\rm Tl})=(69 \pm 111)\times 10^{-12}e~{\rm fm}^{3} \ .
\end{equation}

In Table \ref{tab:finaledm} we present our final
results for the atomic EDMs induced by the Schiff moments $S$
alongside the values of previous calculations/estimates.

The limit on the Schiff moment of $^{129}$Xe
(Eq.~(\ref{eq:Xelimit}), Table \ref{tab:finaledm})
is
\begin{equation}
S(^{129}{\rm Xe})=(184 \pm 868 \pm 26)\times 10^{-12}e~{\rm fm}^{3} \ .
\end{equation}

Our result for Hg places new constraints on the fundamental
$P,T$-violating parameters.
Comparing (\ref{eq:Hglimit}) and (\ref{eq:hgS}) we obtain
for the limit on the nuclear Schiff moment:
\begin{equation}
\label{eq:schifflim}
S(^{199}{\rm Hg})=(3.8 \pm 1.8 \pm 1.4)\times 10^{-12}e~{\rm fm}^{3} \ .
\end{equation}
Comparison of the calculated and measured values of the
ionization potentials and polarizabilities
as well as a comparison of the calculations in two
completely different approximations (in $V^{N}$ and
$V^{N-2}$) indicate that the error of the atomic
calculations probably doesn't exceed 20\%.

In general, the Schiff moment can be induced from a number of
$P,T$-violating mechanisms:
due to a permanent EDM of an unpaired nucleon
or due to the $P,T$-violating nucleon-nucleon interaction.
Mercury has an unpaired neutron, and so in the shell model
its Schiff moment can arise due to the EDM of the neutron.
The contribution to the Schiff moment of Hg from a proton
EDM can be estimated by comparing the experimental value
of the magnetic moment of Hg with that estimated by the
nuclear shell model (this allows us to estimate the contribution
of proton configurations with unpaired spin);
see Ref. \cite{dzuba85}.
The dominant mechanism for the production of a Schiff moment
is the $P,T$-odd nucleon-nucleon interaction $\eta$
\cite{sushkov84}.
Due to this mechanism,
the $P,T$-odd field of the unpaired neutron excites core
protons which contribute to the Schiff moment
\cite{flambaum85i}.
The magnitude of this interaction is characterized by the
dimensionless constant $\eta _{np}$.
In Table \ref{tab:limits} we present the limits
on these parameters extracted from Hg. These are compared in
the table with the best limits from other experiments.
Also presented are limits on $P,T$-violating parameters at
the more fundamental level, the $P,T$-odd pion-nucleon
coupling constant ${\bar{g}_{\pi NN}}^{0}$,
the $P,T$-odd QCD phase $\bar{\theta}$,
and the chromoelectric dipole moments (CEDMs) and EDMs of quarks.
We see here that the limits extracted from the Hg measurement
are stronger than those extracted from direct neutron EDM experiments.

\section*{Acknowledgments}

We would like to thank A.N. Petrov and A.V. Titov for valuable
discussions.
This work was supported by the Australian Research Council.
MK is grateful to UNSW for hospitality and acknowledges support from RFBR,
grant No 02-02-16387.


\begin{table}
\caption{Ionization potentials of Xe, Rn, Hg, Ra, and Pu.
The calculated values are obtained from the Hartree-Fock
approximation in the $V^N$ potential. The experimental values are
presented in the last column. Units: ${\rm cm}^{-1}$.}
\label{tab:ip}
\begin{tabular}{ldd}
Atom & HF & Exp. \tablenotemark[1] \\
\hline
Xe & 96525 & 97834.4 \\
Rn & 84285 & 86692.5 \\
Hg & 71996 & 84184.1 \\
Ra & 36485 & 42577.35 \\
Pu & 41463 & 48890(200) \tablenotemark[2] \\
\end{tabular}
\tablenotetext[1]{From Ref. \cite{moore} unless otherwise stated.}
\tablenotetext[2]{Ref. \cite{blaise}.}
\end{table}
\begin{table}
\caption{Scalar polarizabilities of Xe, Rn, Hg, Ra, and Pu
obtained in HF and TDHF approximations in $V^{N}$.
Experimental values are presented in the last column.
(a.u.)}
\label{tab:alphavn}
\begin{tabular}{lddd}
Atom & HF & TDHF & Exp. \tablenotemark[1] \\
\hline
Xe & 26.87 & 26.97 & 27.06 \\
Rn & 34.42 & 35.00 &  \\
Hg & 40.91 & 44.92 & 34 \\
Ra & 204.2 & 297.0 &  \\
Pu & 147.3 & 201.3 &  \\
\end{tabular}
\tablenotetext[1]{Ref. \cite{radtsig}}
\end{table}
\begin{table}
\caption{Electric dipole moments $d$ induced in
Xe, Rn, Hg, Ra, and Pu by the nuclear Schiff moment $S$.
We present results obtained using the HF
and TDHF approximations
in the $V^{N}$ potential.
Units: $10^{-17}(S/(e~{\rm fm}^{3}))e~{\rm cm}$~.}
\label{tab:dvn}
\begin{tabular}{ldd}
Atom & HF & TDHF \\
\hline
Xe & 0.289 & 0.378 \\
Rn & 2.47 & 3.33 \\
Hg & -1.19 & -2.97 \\
Ra & -1.85 & -8.23 \\
Pu & -2.66 & -10.9 \\
\end{tabular}
\end{table}
\begin{table}
\caption{Binding energies of low states of Hg and Ra
calculated in the $V^{N-2}$ approximation.
The removal energy for both $6s$ electrons for Hg and both
$7s$ electrons for Ra is presented
in the first row of each respective atom.
Energies of excited states are presented relative to the ground state.
Units: ${\rm cm}^{-1}$.}
\label{tab:ipn-2}
\begin{tabular}{lllddd}
Atom & State & & CI & +MBPT & Exp. \tablenotemark[1] \\
\hline
Hg & $6s^{2}$ & $^{1}S_{0}$ & 207659  & 240912 & 235464 \\
   & $6s6p$ & $^{3}P_{0}$ & 29336 & 40012 & 37645.080 \\
   &  & $^{3}P_{1}$ & 31009 & 41753 & 39412.300 \\
   &  & $^{3}P_{2}$ & 34794 & 46776 & 44042.977 \\
   &  & $^{1}P_{1}$ & 46142 & 55395 & 54068.781 \\
   & $6s7s$ & $^{3}S_{1}$ & 51460 & 65044 & 62350.456 \\
   &  & $^{1}S_{0}$ & 54187 & 67090 & 63928.243 \\
Ra & $7s^{2}$ & $^{1}S_{0}$ & 115318 & 127248 & 124419.66 \\
   & $7s6d$ & $^{3}D_{1}$ & 15910 & 14012 & 13715.85 \\
   & & $^{3}D_{2}$ & 16067 & 14465 & 13993.97 \\
   & & $^{3}D_{3}$ & 16625 & 15921 & 14707.35 \\
   & $7s7p$ & $^{3}P_{0}$ & 10424 & 14268 & 13078.44 \\
   & & $^{3}P_{1}$ & 11289 & 15159 & 13999.38 \\
   & & $^{3}P_{2}$ & 13535 & 17937 & 16688.54 \\
   & & $^{1}P_{1}$ & 18835 & 21663 & 20715.71 \\
\end{tabular}
\tablenotetext[1]{Ref. \cite{moore}.}
\end{table}
\begin{table}
\caption{The scalar polarizabilities $\alpha$ for
Hg and Ra calculated in the potential $V^{N-2}$.
The valence and core contributions are
separated into different rows and their sum is presented.
The HF result, with
CI included into the valence orbital contribution, is presented
in the first column of results. In the next columns the Brueckner
and TDHF contributions are added. (a.u.)}
\label{tab:alphavn-2}
\begin{tabular}{llddd}
Atom & Contribution & CI & + MBPT & + TDHF \\
\hline
Hg & $6s^{2}$ & 46.54 & 36.06 & 25.69 \\
& core & 8.05 & 8.05 & 7.30 \\
& sum & 54.59 & 44.11 & 32.99 \\
Ra & $7s^{2}$ & 321.4 & 260.7 & 218.2 \\
& core & 13.6 & 13.6 & 11.8 \\
& sum & 335.0 & 274.3 & 229.9 \\
\end{tabular}
\end{table}
\begin{table}
\caption{The EDMs induced in Hg and Ra calculated
in the $V^{N-2}$ potential.
The valence and core contributions are
separated into different rows and their sum is presented.
The HF result, with
CI included into the valence orbital contribution, is presented
in the first column of results. In the next columns the Brueckner
and TDHF contributions are added.
Units: $10^{-17}(S/(e~{\rm fm}^{3}))e~{\rm cm}$~.}
\label{tab:dvn-2}
\begin{tabular}{llddd}
Atom & Contribution & CI & + MBPT & + TDHF \\
\hline
Hg & $6s^{2}$ & -1.90 & -2.77 & -3.04 \\
& core & 0.63 & 0.63 & 0.34 \\
& sum & -1.26 & -2.14 & -2.70 \\
Ra & $7s^{2}$ & -4.03 & -6.08 & -10.10 \\
& core & 1.62 & 1.62 & 1.40 \\
& sum & -2.41 & -4.46 & -8.70 \\
\end{tabular}
\end{table}
\begin{table}
\caption{Final results for the atomic EDMs induced by the
nuclear Schiff moment. These are compared with the simple
calculations/estimates of previous works.
Units: $10^{-17}(S/(e~{\rm fm}^{3}))e~{\rm cm}$~.}
\label{tab:finaledm}
\begin{tabular}{ldd}
Atom & Other works & This work \\
\hline
Xe & 0.27 \tablenotemark[1] & 0.38 \\
Rn & 2.0 \tablenotemark[2] & 3.3 \\
Hg & -4.0 \tablenotemark[3] & -2.8 \\
Ra & -7.0 \tablenotemark[2] & -8.5 \\
Pu & -10 \tablenotemark[4] & -11 \\
\end{tabular}
\tablenotetext[1]{Ref. \cite{dzuba85}.}
\tablenotetext[2]{Ref. \cite{spevak97}.}
\tablenotetext[3]{Ref. \cite{flambaum85i}.}
\tablenotetext[4]{Ref. \cite{flambaum02}.}
\end{table}
\begin{table}
\caption{Limits on $P,T$-violating parameters in the hadronic
sector extracted from $^{199}$Hg (Eq.~(\ref{eq:schifflim}))
compared with the best limits from
other experiments. Errors are experimental.
Some relevant theoretical works are presented in the last column.}
\label{tab:limits}
\begin{tabular}{lllll}

$P,T$-violating term & Value & System & Exp. & Theory \\
\hline

neutron EDM $d_{\rm n}$ & $(17\pm 8\pm 6)\times 10^{-26}~e~{\rm cm}$
& $^{199}{\rm Hg}$ & \cite{romalis01} & \cite{sandars67,khriplovich} \\

 & $(1.9\pm 5.4)\times 10^{-26}~e~{\rm cm}$ & neutron & \cite{ILL} & \\

 & $(2.6\pm 4.0\pm 1.6)\times 10^{-26}~e~{\rm cm}$ & neutron &
\cite{PNPI} & \\

& & & & \\

proton EDM $d_{\rm p}$ & $(1.7\pm 0.8\pm 0.6)\times 10^{-24}~e~{\rm cm}$
& $^{199}{\rm Hg}$ & \cite{romalis01} &
\cite{sandars67,khriplovich,dzuba85} \\

& $(17\pm 28)\times 10^{-24}~e~{\rm cm}$ & TlF &
\cite{tlfedm} & \cite{sandars67,petrov02} \\

& & & & \\

$\eta _{\rm np}i\frac{G}{\sqrt{2}}\bar{p}{p}\bar{n}\gamma _{5}{n}$&
$\eta _{\rm np}=(2.7\pm 1.3\pm 1.0)\times 10^{-4}$ &
$^{199}{\rm Hg}$ & \cite{romalis01} &
\cite{flambaum85i} \\

& & & & \\

${\bar{g}_{\pi NN}}^{0}$ &
$(3.0\pm 1.4\pm 1.1)\times 10^{-12}$ & $^{199}{\rm Hg}$ &
\cite{romalis01} & \cite{khriplovich} \\

& & & & \\

QCD phase $\bar{\theta}$ &
$(1.1\pm 0.5\pm 0.4)\times 10^{-10}$ &
$^{199}{\rm Hg}$ & \cite{romalis01} & \cite{crewther79,khriplovich} \\

& $(1.6\pm 4.5)\times 10^{-10}$ & neutron & \cite{ILL} &
\cite{pospelov99} \\

& $(2.2\pm 3.3\pm 1.3)\times 10^{-10}$ & neutron & \cite{PNPI} &
\cite{pospelov99} \\

& & & & \\

CEDMs $\tilde{d}$ and &

$e(\tilde{d}_{\rm d}-\tilde{d}_{\rm u})
=(1.5\pm 0.7\pm 0.6)\times 10^{-26}~e~{\rm cm}$
& $^{199}{\rm Hg}$ & \cite{romalis01} & \cite{pospelov} \\

EDMs $d$ of quarks &
$e(\tilde{d}_{\rm d}+0.5\tilde{d}_{\rm u})+1.3d_{\rm d}-0.3d_{\rm u}$
& & & \\

& $\qquad
=(3.5\pm 9.8)\times 10^{-26}~e~{\rm cm}$ & neutron &
\cite{ILL} & \cite{pospelov01} \\

& $\qquad
=(4.7\pm 7.3\pm 2.9)\times 10^{-26}~e~{\rm cm}$ & neutron &
\cite{PNPI} & \cite{pospelov01}\\
\end{tabular}
\end{table}


\begin{thebibliography}{20}

\bibitem{romalis01}

M.V. Romalis, W.C. Griffith, J.P. Jacobs, and E.N. Fortson,
Phys. Rev. Lett. {\bf 12}, 2505 (2001).

\bibitem{schiff63}

L.I. Schiff, Phys. Rev. {\bf 132}, 2194 (1963).

\bibitem{khriplovich}

I.B. Khriplovich and S.K. Lamoreaux,
{\it CP Violation Without Strangeness}
(Springer, Berlin, 1997).

\bibitem{fortson83}

E.N. Fortson, Bull. Am. Phys. Soc. {\bf 28}, 1321 (1983).

\bibitem{flambaum85}

V.V. Flambaum and I.B. Khriplovich,
Zh. \'{E}ksp. Teor. Fiz. {\bf 89}, 1505 (1985)
[JETP {\bf 62}, 872 (1985)].

\bibitem{flambaum85i}

V.V. Flambaum, I.B. Khriplovich, and O.P. Sushkov,
Phys. Lett. {\bf 162B}, 213 (1985);
Nucl. Phys. {\bf A449}, 750 (1986).

\bibitem{mp}

A.-M. M{\aa}rtensson-Pendrill, Phys. Rev. Lett. {\bf 54}, 1153 (1985).

\bibitem{rosenberry01}

M.A. Rosenberry and T.E. Chupp,
Phys. Rev. Lett. {\bf 86}, 22 (2001).

\bibitem{dzuba85}

V.A. Dzuba, V.V. Flambaum, and P.G. Silvestrov,
Phys. Lett. {\bf 154B}, 93 (1985).

\bibitem{spevak97}

V. Spevak, N. Auerbach, and V.V. Flambaum,
Phys. Rev. C {\bf 56}, 1357 (1997).

\bibitem{engel00}

J. Engel, J.L. Friar, and A.C. Hayes,
Phys. Rev. C {\bf 61}, 035502 (2000).

\bibitem{chupp}

T.E. Chupp, talk at ITAMP Workshop:
{\it Tests of Fundamental Symmetries in Atoms and
Molecules}, 29 Nov. - 1 Dec., 2001.

\bibitem{flambaum02}

V.V. Flambaum and J.S.M. Ginges,
Phys. Rev. A {\bf 65}, 032113 (2002).

\bibitem{tlfedm}

D. Cho, K. Sangster, and E.A. Hinds,
Phys. Rev. A {\bf 44}, 2783 (1991).

\bibitem{petrov02}

A.N. Petrov {\it et al},
Phys. Rev. Lett. {\bf 88}, 073001 (2002).

\bibitem{dzuba96}

V.A. Dzuba, V.V. Flambaum, and M.G. Kozlov,
Pis'ma Zh. \'{E}ksp. Teor. Fiz. {\bf 63}, 844 (1996)
[JETP Lett. {\bf 63}, 882 (1996)];
Phys. Rev. A {\bf 54}, 3948 (1996).

\bibitem{dzuba98}

V.A. Dzuba, M.G. Kozlov, S.G. Porsev,
and V.V. Flambaum,
Zh. \'{E}ksp. Teor. Fiz.{\bf 114}, 1636 (1998)
[JETP {\bf 87}, 885 (1998)].

\bibitem{dzuba87}

V.A. Dzuba, V.V. Flambaum, P.G. Silvestrov, and O.P. Sushkov,
J. Phys. B {\bf 20}, 1399 (1987).

\bibitem{bogdanovich83}

P. Bogdanovich and G. \^{Z}ukauskas,
Sov. Phys. Collect. {\bf 23}, 13 (1983).

\bibitem{moore}

C.E. Moore, {\it Atomic Energy Levels},
Natl. Bur. Stand. (U.S.) Circ. No. 467
(NBS, Washington, D.C., 1958).

\bibitem{blaise}

J. Blaise and J.-F. Wyart,
{\it Energy Levels and Atomic Spectra of Actinides},
International Tables of Selected Constants 20
(TIC, Paris, 1992).

\bibitem{radtsig}

A.A. Radtsig and B.M. Smirnov,
{\it Reference Data on Atoms, Molecules, and Ions}
(Springer, Berlin, 1985).

\bibitem{KTMS97}
M. G. Kozlov, A. V. Titov, N. S. Mosyagin, and P. V. Souchko,
Phys. Rev. A {\bf 56}, R3326 (2001).

\bibitem{sushkov84}

O.P. Sushkov, V.V. Flambaum, and I.B. Khriplovich,
Zh. Exp. Teor. Fiz. {\bf 87}, 1521 (1984)
[Sov. Phys. JETP {\bf 60}, 873 (1984)].

\bibitem{sandars67}

P.G.H. Sandars, Phys. Rev. Lett. {\bf 19}, 1396 (1967).

\bibitem{crewther79}

R.J. Crewther, P. Di Vecchia, G. Veneziano, and E. Witten,
Phys. Lett. {\bf 88B}, 123 (1979).

\bibitem{ILL}

P.G. Harris {\it et al}, Phys. Rev. Lett. {\bf 82}, 904 (1999).

\bibitem{PNPI}

I.S. Altarev {\it et al}, Phys. Atom. Nucl. {\bf 59}, 1152 (1996).

\bibitem{pospelov99}

M. Pospelov and A. Ritz, Phys. Rev. Lett. {\bf 83}, 2526 (1999).

\bibitem{pospelov}

M. Pospelov, hep-ph/0109044.

\bibitem{pospelov01}

M. Pospelov and A. Ritz, Phys. Rev. D {\bf 63}, 073015 (2001).

\end{thebibliography}
\end{document}